\documentclass[aps,prl,floatfix,twocolumn,showpacs,reprint,superscriptaddress]{revtex4-1}
\usepackage{amsfonts,stmaryrd}
\usepackage{mathrsfs}
\usepackage{amsmath}
\usepackage{color}
\usepackage{graphicx}
\usepackage{bm}
\usepackage{amssymb}
\usepackage{xspace}
\usepackage{epstopdf}
\usepackage{dcolumn}
\usepackage{longtable}
\usepackage{multirow}
\usepackage[colorlinks=true, letterpaper=true, pdfstartview=FitV, linkcolor=blue, citecolor=blue, urlcolor=blue]{hyperref}

\begin{document}

\preprint{APS/123-QED}

\title{Intrinsic Magnetoconductivity of Non-magnetic Metals}

\author{Yang Gao}
\affiliation{Department of Physics, The University of Texas at Austin, Austin, Texas 78712, USA}

\author{Shengyuan A. Yang}
\affiliation{Research Laboratory for Quantum Materials, Singapore University of Technology and Design, Singapore 487372, Singapore}

\author{Qian Niu}
\affiliation{Department of Physics, The University of Texas at Austin, Austin, Texas 78712, USA}
\affiliation{International Center for Quantum Materials, Peking University, Beijing 100871, China}

\date{\today}

\begin{abstract}
We present a comprehensive study of magnetoconductivity for general three-dimensional non-magnetic metals within the Berry-curvature-corrected semiclassical and Boltzmann framework. We find a new contribution, which is intrinsic in the sense that its ratio to the zero-magnetic-field conductivity is fully determined by the intrinsic band properties, independent of the transport relaxation time, showing a clear violation of Kohler's rule. Remarkably, this contribution can generally be positive for the longitudinal configuration, providing a new mechanism for the appearance of positive magnetoconductivity under longitudinal configuration besides the chiral anomaly effect.
\begin{description}
\item[PACS numbers]72.15.-v, 72.10.Bg, 72.15.Gd

\end{description}
\end{abstract}


\maketitle

Magnetoconductivity, the field-dependent part of the diagonal component in the conductivity tensor: $\delta\sigma(B)\equiv\sigma(B)-\sigma_0$, where $\sigma_0$ is the conductivity under zero magnetic field, has long been a focus in solid-state physics, due to the fascinating complexity in its behavior and the rich information it could offer about the underlying electronic dynamics~\cite{Pippard,Mermin}. Depending on the relative orientation between the current flow and the magnetic field, the magnetoconductivity can be differentiated as transverse or longitudinal. While the magnetoconductivity under transverse configuration ($\bm E\perp \bm B$) can be naturally expected from the Lorentz force, the appearance of the magnetoconductivity under parallel configuration ($\bm E\parallel \bm B$) is a bit surprising because at first look one expects that the electron's motion along the $B$-field should not be affected. Indeed, in the simple Drude-Sommerfeld model for metals~\cite{Mermin}, the magnetoconductivity under parallel configuration vanishes identically.

In the semiclassical regime with small $B$-field, one possible mechanism for the magnetoconductivity under parallel configuration was identified in Ref.~\onlinecite{Pal2010}, arising due to certain special Fermi surface anisotropy. Recently, the interest in magnetotransport was further fueled by the discovery of topological semimetals~\cite{Wan2011,Mura2007,Young2012,BJYang2014}. Particularly, using semiclassical approach, it was predicted that the chiral anomaly effect~\cite{NN1983} associated with the topological band-crossing points can lead to a large positive magnetoconductivity under parallel configuration~\cite{Son2013}. This effect generates great interest in the magneto-transport studies~\cite{Franz2013,JHZhou2013,Burkov2014,Burkov2015,PALee2015,Adam2015,HChen2015,Shen2016}. Experimentally, such positive signal was indeed observed in a number of doped topological semimetal candidates, and has been interpreted as compelling evidence for their topological band structures~\cite{Kim2013,Xiong2015,Huang2015,Reis2016,Zhang2016,Hli2015,Czli2015}.

For non-magnetic metals, due to the constraint of time reversal symmetry and Onsager's relation, $\delta\sigma(B)=\delta\sigma(-B)$, the leading order magnetoconductivity is of $B^2$, hence its theoretical formulation necessarily requires semiclassical equations of motion that are accurate to second order.
However, previous theories (including Refs.~\onlinecite{Pal2010} and \onlinecite{Son2013}) are based on the semiclassical equations of only first-order accuracy, hence the obtained results are not complete. One naturally wonders: is there any important contribution missing from the picture? Meanwhile, peculiar positive magnetoconductivity under parallel configuration appears in recent experiments on several metallic materials~\cite{Luo2016,ZAXu2016,Yli2016}. These materials are not topological semimetals, and the possibility of anisotropic Fermi surface contribution is also ruled out in experiment, clearly pointing to the existence of new contributions missing in the previous theory.

Here, based on the recently developed semiclassical theory with second-order accuracy~\cite{Yang2014,Yang2015}, we formulate a theory of magnetoconductivities under transverse and parallel configurations of non-magnetic metals in the semiclassical regime. We find that besides the previously obtained contributions, there is
a new contribution to the magnetoconductivity that is linear in the transport relaxation time. We name it the intrinsic magnetoconductivity because its ratio to $\sigma_0$ is independent of scattering, only consisting of intrinsic band quantities including the Berry curvature, orbital magnetic moment and so on. This intrinsic magnetoconductivity is generally nonzero regardless of the Fermi surface geometry, and it offers a new mechanism for the violation of Kohler's rule. Importantly, in systems without topological band-crossings and strong surface Fermi anisotropy, this contribution dominates the magnetoconductivity under parallel configuration, and furthermore, its sign could be positive, offering a new mechanism for positive magnetoconductivity under parallel configuration besides the chiral anomaly.

In the semiclassical theory, the Bloch electron dynamics is described by tracing the electron wave-packet center $(\bm r_c,\bm k_c)$ in the phase space~\cite{Xiao2010}. Assuming the simple case where the Fermi level insects with a single band, the electric current can be expressed as
\begin{equation}\label{eq_cur}
\bm j=-\int {d^3 k\over (2\pi)^3} \mathcal{D} \dot{\bm r} f\,,
\end{equation}
where $\mathcal{D}$ is a correction factor for the density of states~\cite{Xiao2005}, and $f$ is the distribution function. Here and hereafter, we set $e=\hbar=1$ and drop the subscript $c$ for the wave-packet coordinates.

We consider the semiclassical regime with small external fields, such that $\omega_c\tau\ll 1$ where $\omega_c$ is the cyclotron frequency and $\tau$ is the transport relaxation time. As we mentioned, for non-magnetic metals, the leading order contribution in the magnetoconductivity $\delta\sigma(B)$ is of $B^2$. This means that we have to keep the third-order terms in the current (Eq.~(\ref{eq_cur})) that are $\propto EB^2$, where $E$ is the electric field strength. To this end, as we will show in a while, the following second-order semiclassical equations of motion are sufficient~\cite{Yang2014}:
\begin{align}
\label{eq_rdot} \dot{\bm r}&=\partial_{\bm k}\tilde{\varepsilon}-\dot{\bm k}\times \tilde{\bm \Omega}\,,\\
\label{eq_kdot} \dot{\bm k}&=-\bm E-\dot{\bm r}\times \bm B\,.
\end{align}
Here $\tilde{\varepsilon}$ is the band energy including field-corrections up to second order~\cite{Yang2015}, $\tilde{\bm \Omega}$ is the modified Berry curvature including first-order field corrections. Due to the non-canonical structure of the equations of motion, the correction factor $\mathcal{D}$ in Eq.~(\ref{eq_cur}) takes the form $\mathcal{D}=1+\bm B\cdot \tilde{\bm \Omega}$~\cite{Yang2014}.

The remaining factor in in Eq.(\ref{eq_cur}), i.e. the distribution function $f$, is typically solved from the Boltzmann equation. For a homogeneous system at steady state, it reads
\begin{equation}\label{eq_boltz}
\dot{\bm k}\cdot {\partial f\over \partial \bm k}=\left . {df\over dt}\right |_\text{collison}\,.
\end{equation}
Here $\dot{\bm k}$ can be substituted from the Eq.~(\ref{eq_kdot}), and the collision integral on the right hand side describes the relaxation due to the various scattering processes in the system. To proceed analytically, we take the relaxation time approximation such that the right hand side of Eq.(\ref{eq_boltz}) becomes $-(f-f_0)/\tau$, where $f_0$ is the equilibrium Fermi distribution and relaxation process is characterized by a single transport relaxation time $\tau$. Note that in Eq.(\ref{eq_boltz}), the argument of the equilibrium distribution function $f_0$ must be the band energy including the magnetic field corrections, such that it guarantees a vanishing current at $E=0$. Then the solution of the Boltzmann equation can be obtained as
\begin{equation}\label{eq_dis}
f=\sum_{m=0}^\infty(-\tau \dot{\bm k}\cdot \bm \partial_{\bm k})^m f_0(\tilde{\varepsilon})\,.
\end{equation}

Eqs.~(\ref{eq_rdot}), (\ref{eq_kdot}), and (\ref{eq_dis}) offer all the necessary and sufficient ingredients in evaluating the current in Eq.~(\ref{eq_cur}) to third order in external fields. To see this, one notes that since the equilibrium distribution $f_0$ does not contribute to the current, the factor $f$ in Eq.~(\ref{eq_cur}) is at least of first order ($\dot{\bm k}$ in Eq.~(\ref{eq_dis}) is at least of first order according to Eq.~(\ref{eq_kdot})). Hence each of the other two factors $\mathcal{D}$ and $\dot{\bm r}$ in Eq.~(\ref{eq_cur}) only needs to be accurate to second order. And this is why it is sufficient to have the Berry curvature $\tilde{\bm \Omega}$ in both $\mathcal{D}$ and $\dot{\bm r}$ to be corrected to first order and $\tilde{\varepsilon}$ in $\dot{\bm r}$ and $f_0$ to be corrected up to second order.

Straightforward substitution and calculation yields the following contributions to the current up to third order~\cite{suppl}:
\begin{widetext}
\begin{align}\label{eq_j}
\bm j^{(a)}=-\tau^3\int {d^3 k\over (2\pi)^3}  \bm v_0 (\bm v_0\times \bm B\cdot \bm \partial_{\bm k})^2(\bm E\cdot \bm \partial_{\bm k}) f_0(\varepsilon_0)-\tau \int {d^3 k\over (2\pi)^3} [\tilde{\bm v}+(\tilde{\bm v}\cdot\tilde{\bm \Omega}) \bm B][\bm E+(\bm E\times \tilde{\bm \Omega})\times \bm B]\cdot \bm \partial_{\bm k} f_0(\tilde{\varepsilon})\,.
\end{align}
\end{widetext}
Here we group the various terms into two compact terms according to their $\tau$ dependence, $\varepsilon_0$ is the unperturbed band energy, $\bm v_0=\bm \partial_{\bm k}\varepsilon_0$ and $\tilde{\bm v}=\bm \partial_{\bm k}\tilde{\varepsilon}$ are the band velocities for the unperturbed and perturbed band dispersions, respectively. Both terms here are Fermi surface contributions, i.e. carrying the derivative of the Fermi distribution function. We observe that the first term in Eq.~(\ref{eq_j}) just recovers the contribution from Fermi surface anisotropy identified in Ref.~\onlinecite{Pal2010}, while the second term is new.

In addition, at second order, external fields induce a shift $\delta\mu$ in the chemical potential (the linear-order correction vanishes as in usual first order transport theory), which could lead to additional contribution to the current. Since $\delta\sigma_{xx}\sim B^2$, there are only two possible field-dependence in $\delta\mu$ that could contribute, $\delta\mu\sim EB$ or $\sim B^2$. The first possibility happens only with special band-crossings, i.e., in the chiral anomaly effect for doped Weyl semimetals~\cite{Son2013}. In that case, the parallel $E$ and $B$ fields pumps charges between a pair of Weyl points, shifting the chemical potential around each Weyl point in opposite ways. The current contributed from each Weyl point can be expressed as: $\bm j^{\text{CA}}=\int \frac{d^3k}{(2\pi)^3}\bm B(\bm v_0\cdot\bm \Omega_0)\delta\mu f_0'$~\cite{Son2013}. Here $\bm\Omega_0$ is the usual (unperturbed) Berry curvature, and $\delta\mu\propto \tau_v\chi \bm E\cdot\bm B$ where $\chi$ is the chirality of the Weyl point and  $\tau_v$ is the intervalley scattering time that sustains the electron population imbalance between the pair of Weyl points.

The other possible contribution from $\delta\mu\sim B^2$ is more general, regardless of the band topology. It yields the following current that contributes to the magnetoconductivity:
\begin{equation}\label{eq_jadd}
\bm j^{(b)}=\tau \delta\mu\int {d^3 k\over (2\pi)^3} \bm v_0 ( \bm v_0\cdot\bm E) f_0''\,.
\end{equation}
The shift $\delta\mu$ can be fixed from the particle number conservation. For example, when the conduction band is separated from the valence band, the electron number $n$ in the conduction band should be conserved. Under $B$-field, the shift $\delta\mu$ compensates the field correction of the band dispersion to ensure that $n=\int \frac{d^3k}{(2\pi)^3} \mathcal{D} f_0(\tilde{\varepsilon}-\mu-\delta\mu)$ is a constant, from which we can solve out $\delta\mu$. Note that without $E$-field dependence, this shift $\delta\mu$ is an equilibrium property and is the same across the Brillouin zone.

This completes our general analysis for all possible contributions to the magnetoconductivity.  Since we did not specify the directions of the fields, the result holds for magnetoconductivities under both the longitudinal and transverse  configurations.

For conventional metals (without chiral anomaly), the current up to third order is given by $\bm j=\bm j^{(a)}+\bm j^{(b)}$. One observes that the current at third order only has $\tau$-dependence as $\sim\tau$ or $\sim\tau^3$. This is because that $\tau$ is coupled with $\dot{\bm k}$ in
Eq.~(\ref{eq_dis}), its power cannot exceed $\tau^3$. Meanwhile, the term proportional to $\tau^2$ is given by
$-\tau^2 \int{d^3 k\over (2\pi)^3} [\tilde{\bm v}+(\bm v_0\cdot \bm \Omega_0)\bm B](\bm v_0\times \bm B\cdot \bm \partial_{\bm k})(\bm E\cdot \bm \partial_{\bm k}) f(\tilde{\varepsilon})$, which is odd under time-reversal operation hence vanishes identically.

As we have mentioned, the first term in $\bm j^{(a)}$ recovers the result obtained in Ref.~\onlinecite{Pal2010}. It can be shown that its contribution is always negative for magnetoconductivity under transverse condiguration. And for longitudinal configuration, its contribution is nonzero only when the Fermi surface has special anisotropy~\cite{Pal2010}.

Besides recovering the previously known contributions, most importantly, we discover new contributions, including $\bm j^{(b)}$ and the second term in $\bm j^{(a)}$. These terms only appear when using a complete second order semiclassical theory, hence they are missing in the previous works. Due to their common $\tau$-linear dependence, we combine the two terms together, and name their resulting contribution to $\delta\sigma$ as the intrinsic magnetoconductivity ($\delta\sigma^\text{int}$), because the ratio $\delta\sigma^\text{int}/\sigma_0$ is independent of $\tau$, consisting entirely of intrinsic band quantities (apart from the $B^2$ factor).

Our result has important implications on Kohler's rule~\cite{Pippard}. Kohler's rule states that the ratio $\delta\sigma/\sigma_0$ depends on the $B$-field through the quantity $\omega_c\tau$. Since $\sigma_0\propto\tau$, samples with different relaxation times can be related by plotting the ratios against a rescaled field $B\sigma_0$: $\delta\sigma/\sigma_0=F(B\sigma_0)$. Kohler's rule can be derived in the first-order semiclassical theory by assuming a single species of charge carriers and a single relaxation time. Any deviation from Kohler's rule is usually interpreted as from factors beyond the semiclassical description or from the presence of multiple types of carriers or multiple scattering times~\cite{McKe1998}. Here, we see that first term in
$\bm j^{(a)}$, the previously obtained contribution from first-order semiclassical theory, indeed obeys Kohler's rule. Denoting its contribution to the magnetoconductivity as the extrinsic one $\delta\sigma^\text{ext}$, we have $\delta\sigma^\text{ext}/\sigma_0\sim (\omega_c\tau)^2$. However, the new intrinsic contribution clearly violates Kohler's rule because $\delta\sigma^\text{int}/\sigma_0$ is independent of $\tau$. Instead, we may regard $\delta\sigma^\text{int}/\sigma_0\sim (\omega_c\tau_b)^2$ by replacing $\tau$ with another intrinsic time scale  $\tau_b=\hbar/\varepsilon$, where $\varepsilon$ is an intrinsic energy scale for the band, such as the band gap and the chemical potential. Because our derivation is still within the semiclassical framework and under the same conditions on carrier type and relaxation time, it represents a new mechanism for the violation of Kohler's rule. Due to their different $\tau$-dependence, one may expect that $\delta\sigma^\text{ext}$ generally dominates over $\delta\sigma^\text{int}$ in clean samples where $\tau$ is large, for which Kohler's rule could hold to a good extent.

Another important consequence of our result is that the intrinsic contribution offers a new origin of the peculiar magnetoconductivity under longitudinal configuration. It becomes the dominant term in the magnetoconductivity under longitudinal configuration for systems with relatively isotropic Fermi surfaces and without Weyl band-crossing points. As illustrated in the following example, this term can be sizable for bands with nonzero Berry curvatures and orbital magnetic moments, and more remarkably, it can in general be positive. Therefore, our theory provides a new mechanism for positive magnetoconductivity under longitudinal configuration in conventional metals.

In the following, we apply our theory to two concrete examples. In the first example, we consider the low-energy model of a doped Weyl semimetal. Near the Fermi energy, Weyl semimetals have isolated Weyl points, each described by a Weyl Hamiltonian: $H=\chi v_F \bm k\cdot \bm \sigma$, where $\bm \sigma$ is the vector of Pauli matrices denoting the two crossing bands, $\chi=\pm 1$ gives the chirality, and $v_F$ is the Fermi velocity. When time reversal symmetry is preserved, there are at least two pairs of Weyl points in the Brillouin zone.

Consider electron-doped case with $\mu>0$, and assume that the magnetic field is along the $z$-direction. With two pairs of Weyl points, we find that the intrinsic magnetoconductivity in the longitudinal and transverse configuration is given by
\begin{equation}\label{eq_Dirac_int}
\delta\sigma_{\parallel}^{\rm int}={2\over 15}\sigma_0(\omega_c\tau_b)^2\,,\qquad \delta\sigma_{\perp}^{\rm int}=-{17\over 30}\sigma_0 (\omega_c\tau_b)^2\,,
\end{equation}
for longitudinal and transverse configurations respectively. Here the zero-magnetic-field conductivity $\sigma_0=2\mu^2\tau/(3\pi^2v_F)$, $\omega_c=v_F^2B/\mu$, and $\tau_b=1/\mu$ is the Fermi time scale. One observes that the intrinsic magnetoconductivity takes different signs for the two configurations, and more importantly, the intrinsic magnetoconductivity under longitudinal configuration can be positive.

In comparison, the extrinsic contribution as from the first term in $\bm j^{(a)}$ is given by
\begin{equation}\label{eq_Dirac_ext}
\delta\sigma_{\parallel}^{\rm ext}=0\,,\qquad \delta\sigma_{\perp}^{\rm ext}=-\sigma_0 (\omega_c\tau)^2\,.
\end{equation}
Here $\delta\sigma_{\parallel}^{\rm ext}$ is zero because the Fermi surface is isotropic. And its contribution to the magnetoconductivity under transverse configuration $\delta\sigma_{\perp}^{\rm ext}$ is nonzero and follows Kohler's rule as expected.

Finally, in Weyl semimetals, there is a positive magnetoconductivity under parallel configuration from the chiral anomaly effect $\delta\sigma_{\parallel}^\text{CA}$. It is interesting to consider the ratio between the intrinsic magnetoconductivity $\delta\sigma_{\parallel}^{\rm int}$ and the chiral anomaly one:
\begin{equation}\label{eq_ratio}
 {\delta\sigma_{\parallel}^{\rm int}\over \delta\sigma_{\parallel}^\text{CA}}={8\over 45} {\tau\over \tau_{v}}\,.
\end{equation}
We observe that this ratio only depends on the ratio between the two relaxation times.
Generally, the intervalley scattering time $\tau_v$ is much larger than $\tau$~\cite{Burkov2015}, so in doped Weyl semimetals, the chiral anomaly contribution is more important than the intrinsic contribution for the positive magnetoconductivity under longitudinal configuration.

When magnetoconductivities under transverse and longitudinal configurations are of different signs, if we change the mutual orientation between $E$-field and $B$-field from parallel to perpendicular configuration, then for each $B$-field strength, there will be a critical angle $\theta_c$ between the two fields at which the magnetoconductivity changes sign. Using the result obtained above, we find that the critical angle for this simple Weyl model is given by
\begin{equation}
\theta_c(B)=\text{arccot}{34+60(\omega_c\tau)^2\over 8+45(\tau_{v}/\tau)}\,.
\end{equation}
The point is that although here the intrinsic contribution to the magnetoconductivity under longitudinal configuration is relatively small, its contribution to the magnetoconductivity under transverse configuration can be important provided that $\tau_b/\tau$ is sizable. Hence the intrinsic contribution must be included in calculating the critical angle.

In the second example, we consider a two-band model without any band-crossing point. It has two valleys in the Brillouin zone connected by the time reversal symmetry
\begin{equation}\label{eg2}
H_\chi=\chi v_F k_x\sigma_x+ v_Fk_y\sigma_y+\left(\Delta+\frac{k_z^2}{2m^*}\right)\sigma_z.
\end{equation}
Here $\chi=\pm 1$ labels the two valleys, $v_F$, $\Delta$, and $m^*$ (assumed to be positive) are model parameters, and the pseudospin $\sigma_i$ here denotes the two-band degree of freedom. The two bands are separated by a gap of $2\Delta$. Such a continuum model can be derived from a lattice model, e.g., a model defined on a 3D lattice consisting of 2D honeycomb lattices AA-stacked along the $z$-direction~\cite{suppl}.

\begin{figure}[t]
\setlength{\abovecaptionskip}{0pt}
\setlength{\belowcaptionskip}{1pt}
\scalebox{0.29}{\includegraphics*{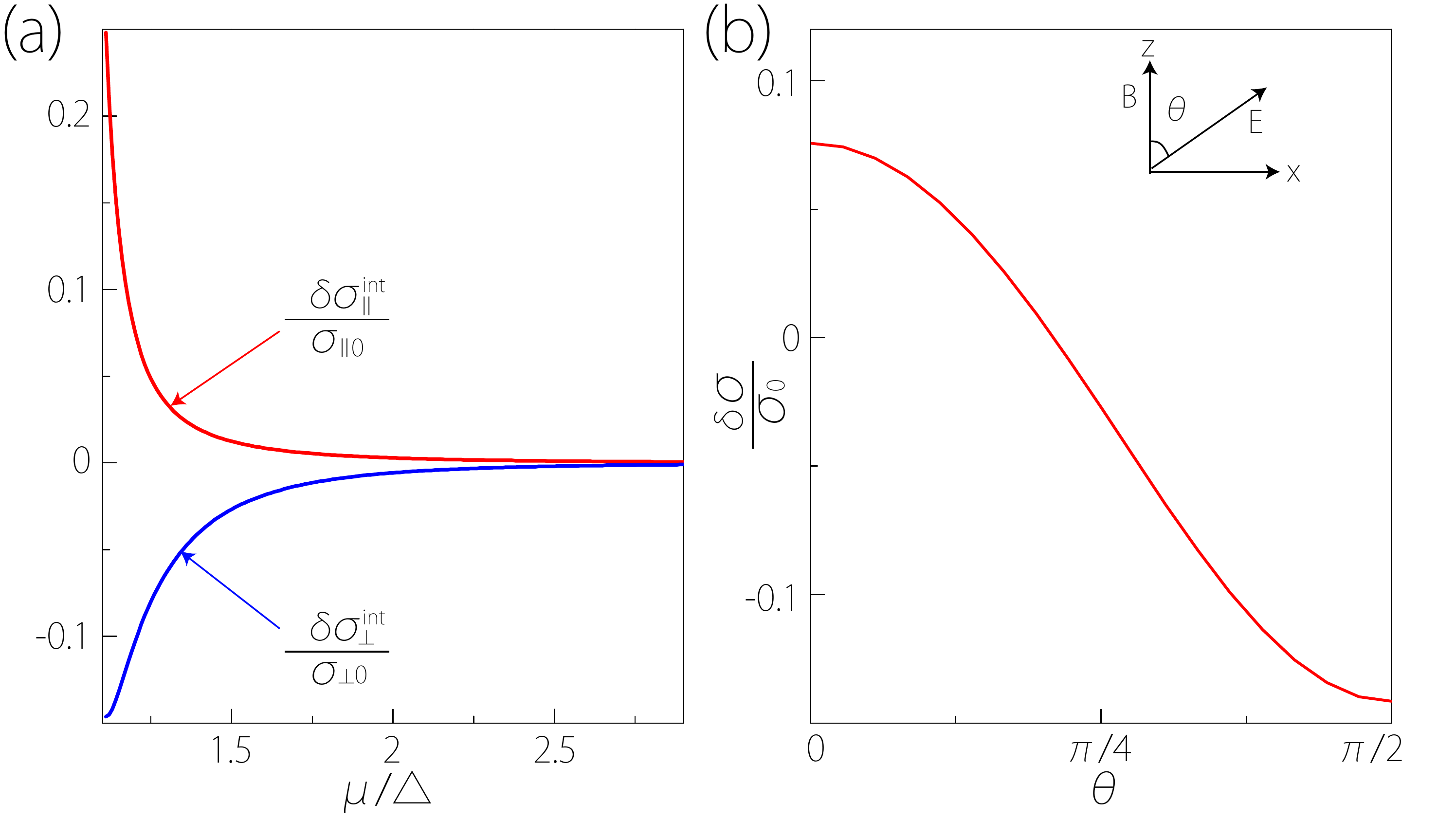}}
\caption{(a) Intrinsic magnetoconductivities for model (\ref{eg2}) versus the chemical potential. Here $\sigma_{\perp 0}$ and $\sigma_{\parallel 0}$ are zero-magnetic-field conductivities under transverse and longitudinal configurations respectively. (b) Ratio between magnetoconductivity $\delta\sigma$
and zero-magnetic-field resistivity $\sigma_0$ versus the angle $\theta$ between $E$ and $B$ fields, as illustrated in the inset. Here the model parameters are chosen as
$B=2$T, $\Delta=50{\rm meV}$, $v_F=9.2\times 10^5 {\rm m/s}$, and $m^*=0.1m_e$ ($m_e$ is the free electron mass). In (b) we take $\mu=60 {\rm meV}$.}
\label{fig_fig1}
\end{figure}

Consider the electron-doped case with $\mu>\Delta$ and take the $B$-field to be along the $z$-direction. For magnetoconductivity under longitudinal configuration, it is clear that there is no chiral anomaly contribution in this model since there is no Weyl point. And because of the axial symmetry of the Fermi surface,
the extrinsic contribution $\delta\sigma_{\parallel}^\text{ext}$ also vanishes~\cite{Pal2010}. The only nonzero contribution here is the intrinsic one, and we find that it gives a positive magnetoconductivity when $\bm E\parallel \bm B$, as illustrated in Fig.~\ref{fig_fig1}(a). From the figure, we observe that the magnitude of the intrinsic magnetoconductivity decreases as $\mu$ increases. This is because that the intrinsic magnetoconductivity is a Fermi surface property that highly depends on the geometric quantities such as Berry curvature and orbital magnetic moment, which are concentrated near the band edge.

In Fig.~\ref{fig_fig1}(b), we plot the magnetoconductivity containing both intrinsic and extrinsic contributions against the mutual orientation of the fields. Here $B$-field is taken to be along the $z$-direction, and $\theta$ is the angle between $E$ and $B$ fields. One observes that the magnetoconductivity gradually changes from positive in the longitudinal case to negative in the transverse configuration. The corresponding magnetoresistivity will change from negative to positive in the process.

To conclude, we have formulated a theory of the magnetoconductivity for non-magnetic metals in the semiclassical regime. We obtain an important new
contribution that is missing in previous theories. This intrinsic contribution provides a new mechanism for the violation of Kohler's rule and for a nonzero magnetoconductivity under longitudinal configuration. Particularly, it dominates the magnetoconductivity under longitudinal configuration in systems without strong Fermi surface anisotropy and topological band-crossings. Furthermore, its value can in general be positive, hence offering a new origin for positive magnetoconductivity besides the chiral anomaly effect. It may play an important role behind the puzzling magnetotransport signal observed in recent experiments on ${\rm TaAs_2}$ and related materials~\cite{Luo2016,ZAXu2016,Yli2016,Wang2016,Yuan2016,ZAXu2016b}.

\begin{acknowledgments}
We acknowledge useful discussions with Hua Chen and Xiao Li, and thank Shan-Shan Wang for help with the figures. QN is supported by NBRPC (No. 2012CB921300 and No. 2013CB921900), and NSFC (No. 91121004). YG is supported by DOE (DE-FG03-02ER45958, Division of Materials Science and Engineering) and Welch Foundation (F-1255). SAY is supported by Singapore MOE Academic Research Fund Tier 2 (MOE2015-T2-2-144).
\end{acknowledgments}

\bibliographystyle{apsrev4-1}

\begin{thebibliography}{100}

\bibitem{Mermin}  N. W. Ashcroft and N. D. Mermin, \emph{Solid State Physics} (Harcourt, Orlando, 1976).
\bibitem{Pippard} A. B. Pippard, \emph{Magnetoresistance in Metals} (Cambridge University Press, Cambridge, 1989).

\bibitem{Pal2010} H. K. Pal and D. L. Maslov, Phys. Rev. B {\bf 81}, 214438 (2010).

\bibitem{Wan2011}  X. Wan, A. M. Turner, A. Vishwanath, and S. Y. Savrasov, Phys. Rev. B {\bf 83}, 205101 (2011).
\bibitem{Mura2007} S. Murakami, New J. Phys. {\bf 9}, 356 (2007).
\bibitem{Young2012}  S. M. Young, S. Zaheer, J. C. Y. Teo, C. L. Kane, E. J. Mele, and A. M. Rappe, Phys. Rev. Lett. {\bf 108}, 140405 (2012).
\bibitem{BJYang2014}  B.-J. Yang and N. Nagaosa, Nat. Commun. {\bf 5}, 4898 (2014).

\bibitem{NN1983}  H. Nielsen and M. Ninomiya, Phys. Lett. B {\bf 130}, 389 (1983).
\bibitem{Son2013} D. T. Son and B. Z. Spivak, Phys. Rev. B {\bf 88}, 104412 (2013).

\bibitem{Franz2013}  M. M. Vazifeh and M. Franz, Phys. Rev. Lett. {\bf 111}, 027201 (2013).
\bibitem{JHZhou2013} J.-H. Zhou, H. Jiang, Q. Niu, and J.-R. Shi, Chin. Phys. Lett. {\bf 30}, 027101 (2013).
\bibitem{Burkov2014} A. A. Burkov, Phys. Rev. Lett. {\bf 113}, 247203 (2014).
\bibitem{Burkov2015} A. A. Burkov, Phys. Rev. B {\bf 91}, 245157 (2015).
\bibitem{PALee2015}  J. C. W. Song, G. Refael, and P. A. Lee, Phys. Rev. B {\bf 92}, 180204(R) (2015).
\bibitem{Adam2015}   N. Ramakrishnan, M. Milletari, and S. Adam, Phys. Rev. B {\bf 92}, 245120 (2015).
\bibitem{HChen2015}  H. Chen, Y. Gao, D. Xiao, A. H. MacDonald, and Q. Niu, arXiv:1511.02557.
\bibitem{Shen2016}   S.-B. Zhang, H.-Z. Lu, and S.-Q. Shen, New J. Phys. {\bf 18}, 053039 (2016).

\bibitem{Kim2013} H.-J. Kim, K.-S. Kim, J.-F. Wang, M. Sasaki, N. Satoh, A. Ohnishi, M. Kitaura, M. yang, and L. Li, Phys. Rev. Lett. {\bf 111}, 246603 (2013).
\bibitem{Xiong2015} J. Xiong, S. K. Kushwaha, T. Liang, J. W. Krizan, M. Hirschberger, W. Wang, R. J. Cava, and N. P. Ong, Science {\bf 350}, 413 (2015).
\bibitem{Huang2015} X. Huang et al., Phys. Rev. X {\bf 5}, 031023 (2015).
\bibitem{Czli2015} C.-Z. Li, L.-X. Wang, H. Liu, J. Wang, Z.-M. Liao, and D.-P. Yu, Nat. Commun. {\bf 6}, 10137 (2015).
\bibitem{Reis2016} R. D. dos Reis, M. O. Ajeesh, N. Kumar, F. Arnold, C. Shekhar, M. Naumann, M. Schmidt, M. Nicklas, and E. Hassinger, New J. Phys. {\bf 18}, 085006 (2016).
\bibitem{Zhang2016} C.-L. Zhang et al., Nat. Commun. {\bf 7}, 10735 (2016).
\bibitem{Hli2015} H. Li, H. He, H.-Z. Lu, H. Zhang, H. Liu, R. Ma, Z. Fan, S.-Q. Shen, and J. Wang, Nat. Commun. {\bf 7}, 10301 (2016).

\bibitem{Luo2016} Y. Luo, R. D. McDonald, R. F. S. Rosa, B. Scott, N. Wakeham, N. J. Ghimire, E. D. Bauer, J. D. Thompson, and F. Ronning, Scientific Reports {\bf 6}, 27294 (2016).
\bibitem{ZAXu2016} Y. Li, Z. Wang, Y. Lu, X. Yang, Z. Shen, F. Sheng, C. Feng, Y. Zheng, and Z.-A. Xu, arXiv:1603.04056.
\bibitem{Yli2016} Y. Li, L. Li, J. Wang, T. Wang, X. Xu, C. Xi, C. Cao, and J. Dai, arXiv:1601.02062.

\bibitem{Yang2014} Y. Gao, S. A. Yang, and Q. Niu, Phys. Rev. Lett. {\bf 112}, 166601 (2014).
\bibitem{Yang2015} Y. Gao, S. A. Yang, and Q. Niu, Phys. Rev. B {\bf 91}, 214405 (2015).
\bibitem{Xiao2010} D. Xiao, M.-C. Chang, and Q. Niu, Rev. Mod. Phys. {\bf 82}, 1959 (2010).
\bibitem{Xiao2005} D. Xiao, J. Shi, and Q. Niu, Phys. Rev. Lett. {\bf 95}, 137204 (2005).

\bibitem{suppl} See Supplemental Materials.

\bibitem{McKe1998} R. H. McKenzie, J. S. Qualls, S. Y. Han, and J. S. Brooks, Phys. Rev. B {\bf 57}, 11854 (1998).

\bibitem{Wang2016} Y. Wang, Q.-H. Yu, P.-J. Guo, K. Liu, and T.-L. Xia, Phys. Rev. B {\bf 94}, 041103(R) (2016).
\bibitem{Yuan2016} Z. Yuan, H. Lu, Y. Liu, J. Wang, and S. Jia, Phys. Rev. B {\bf 93}, 184405 (2016).
\bibitem{ZAXu2016b} Z. Wang, Y. Li, Y. Lu, Z.-X. Shen, F. Sheng, C. Feng, Y. Zheng, and Z.-A. Xu, arXiv:1603.01717.

\end{thebibliography}

\end{document}